\documentstyle[preprint,aps]{revtex}
\begin{document}
\draft
\tighten
\preprint{LA-UR-93-4315}

\widetext
\title{
$\protect\protect\bbox{(j,0)\oplus(0,j)}$ Representation Space:  Dirac-Like
Construct
\footnotemark[1]
\footnotetext[1]{This work was done under the auspices of the U. S.
Department of Energy. The lecture was delivered by D.V.Ahluwalia}
}
\author{D. V. Ahluwalia, M. B. Johnson  and  T. Goldman }
\address{ Los Alamos National Laboratory,
Los Alamos, New Mexico 87545, USA}
\maketitle

\begin{abstract}
This is first of  the two  invited lectures presented at the ``XVII
International School of Theoretical Physics: Standard Model and Beyond' 93.''
The text is essentially based on a recent publication by the present authors
[Phys. Lett. B  316  (1993) 102]. Here we show that the Dirac-like
construct in the $(j,0)\oplus(0,j)$ representation space supports a
Bargmann-Wightman-Wigner-type quantum field theory.

\end{abstract}

\newpage
The class of theories in which  a boson and
antiboson have opposite intrinsic parities is informally known as
``Wigner-type,'' but in view of Wigner's note [1, p. 38]
 that ``much of the'' work in Ref.
\cite{EW} ``was taken from a rather old but unpublished manuscript by V.
Bargmann,
A. S. Wightman and myself,'' we take the liberty of calling this type of theory
a %
Bargmann-Wightman-Wigner-type (BWW-type) quantum field theory. Even though the
generality of BWW's arguments  is remarkable, at present no explicit
construction of a BWW-type quantum field theory is known to exist. Nor has it
been realised that the $(j,0)\oplus(0,j)$ representation for massive particles
is a realisation of the BWW type quantum field theory. In this paper,
we show
that this is the case. We do this by considering the special case of the
$(1,0)\oplus(0,1)$ field and working out explicitly its properties under C, P,
and T.

We begin with a brief review of $(j,0)\oplus(0,j)$ representation space.
In the notation of  Refs. \cite{DVA,LR}, $(j,0)$ and  $(0,j)$
spinors have the following Lorentz transformation properties
\begin{eqnarray}
&&(j,0):\quad \phi_R(\vec p\,)\
=\,\Lambda_R\,\phi_R(\vec 0\,) \,=\,\exp\left(+\,\vec
J\cdot\vec\varphi\,\right)\,\phi_R(\vec 0\,)\quad,\label{r}\\
&&(0,j):\quad\phi_L(\vec p\,)\,
=\,\Lambda_L\,\phi_L(\vec 0\,) \,=\,\exp\left(-\,\vec J\cdot\vec\varphi\,
\right)\,\phi_L(\vec 0\,)\quad. \label{l}
\end{eqnarray}
 The $\vec J$ are the standard $(2j+1)\times(2j+1)$
angular momentum matrices, and $\vec \varphi$ is the boost parameter
defined as
\begin{equation}
\cosh(\varphi\,)
\,=\,\gamma\,=\,{1\over\sqrt{1-v^2}}\,=\,{E\over m},\quad\quad \sinh(
\varphi\,)\,=\,v\gamma\,=\,{| \vec p\, |\over m},\quad\quad\hat\varphi={\vec  p
\over| \vec p\,|},
\end{equation}
with $\vec p$ the three-momentum of the particle. In order to stay as close as
possible to the standard treatments \cite{LR}  of the $(1/2,0)\oplus(0,1/2)$
 Dirac field, we now
introduce
$(j,0)\oplus(0,j)$-spinor in the
generalised canonical representation [2a,2f]

\begin{equation}
\psi(\vec p\,)\,=\,{1\over\sqrt{2}}
\left(
\begin{array}{cc}
\phi_R(\vec p\,)\,+\,\phi_L(\vec p\,)\\
\phi_R(\vec p\,)\,-\,\phi_L(\vec p\,)\\
\end{array}\right)\quad,\label{crs}
\end{equation}
In the  $(j,0)\oplus(0,j)$ representation space, there are $(2j+1)$
``$u_\sigma(\vec p\,)$ spinors'' and $(2j+1)$ ``$v_\sigma(\vec p\,)$ spinors.''
\footnotemark[1]
\footnotetext[1]{It should be noted that at this stage no physical meaning
need be attached to this distinction between  the $u$- and $v$- spinors.
We are simply choosing our basis spinors to span a $2(2j+1)$ representation
space and anticipating a %
certain physical interpretation.
We will soon see that the  $u$- and $v$- spinors transform differently
under the operation of parity {\it and} satisfy two different equations.}

As a consequence of the  transformation properties (1,2),
and the definition (\ref{crs}), these spinors transform as
\begin{equation}
\psi(\vec p\,)\,=\,M(j,\,\vec p\,)\,\psi(\vec 0\,)\,=\,
\left(
\begin{array}{ccc}
\cosh(\vec J\cdot\vec\varphi)&{\,\,} &\sinh(\vec J\cdot\vec\varphi) \\
\sinh(\vec J\cdot\vec\varphi)&{\,\,} &\cosh(\vec J\cdot\vec\varphi)
\end{array}\right)\,\psi(\vec 0\,)\quad.
\label{b}
\end{equation}
If we work in a representation of the $\vec J$ matrices in which $J_z$ is
diagonal, then the rest spinors
$u_\sigma(\vec 0\,)$ and $v_\sigma(\vec 0\,)$,
$\sigma=j,j-1,\cdots,-j$, can be written in the form of
$2(2j+1)$ single-column matrices each with $2(2j+1)$ elements as follows: %
\begin{equation}
u_{+j}(\vec 0\,)\,=\,
\left(
\begin{array}{c}
N(j)\\
0\\
\vdots\\
0
\end{array}\right)
\,,
u_{j-1}(\vec 0\,)\,=\,
\left(
\begin{array}{c}
0\\
N(j)\\
\vdots\\
0
\end{array}\right)
\quad,
\cdots,\quad
v_{-j}(\vec 0\,)\,=\,
\left(
\begin{array}{c}
0\\
\vdots\\
0\\
N(j)
\end{array}\right)\quad.\label{rs}
\end{equation}
For convenience, and so that the rest spinors vanish in the $m\rightarrow 0$
limit, we choose the normalisation factor $N(j)=m^j$.
With this choice of the normalisation,
the spinors $u_\sigma(\vec p\,)$ and
$v_\sigma(\vec p\,)$ are normalised as follows:
\begin{equation}
\overline{u}_\sigma(\vec p\,)\,u_{\sigma'}(\vec
p\,)=m^{2j}\delta_{\sigma\sigma'}\,,\quad
\overline{v}_\sigma(\vec p\,)\,v_{\sigma'}(\vec
p\,)=-m^{2j}\delta_{\sigma\sigma'}\,,\,\, {\mbox{\rm with}}\,\,
\overline{\psi}_\sigma(\vec p\,)\,=\,
{\psi}_\sigma^\dagger(\vec p\,)\,\Gamma^0\quad.
\end{equation}
 Here, $\Gamma^0$ is a block diagonal matrix with $(2j+1)\times(2j+1)$
identity matrix $I$ on the upper left corner and $-I$ on the  bottom right
corner.
The reader would
immediately realise that for $\vec J=\vec\sigma/2$, $\vec\sigma$ being Pauli
matrices,  the boost $M(1/2,\,\vec p\,)$ is identical to the canonical
representation boost \cite{LR}  for the $(1/2,0)\oplus (0,1/2)$
Dirac spinors. While investigating the $C$, $P$, and
T properties, we will consider the
$(1,0)\oplus(0,1)$ field as an example case. Hence, we present the explicit
expressions for the $(1,0)\oplus(0,1)$ spinors
(in the generalised canonical
representation introduced above). The three $u_{0,\pm 1}(\vec p\,)$
and the three $v_{0,\pm 1}(\vec p\,)$ spinors read:
\footnotemark[2]
\footnotetext[2]{For additional details, the reader should refer to Ref.
[2a]. This reference also provides $(j,0)\oplus(0,j)$ spinors for
$j=3/2$ and $j=2$.}
{\footnotesize
\begin{eqnarray}
u_{_ {+1}}( \vec p\,)&\,=\,&
\pmatrix{m+\left[(2p_z^2\,+\,p_{_{+}} p_{_{-}}) / 2(E+m)\right]\cr
                      {p_z p_{_{+}}/{\sqrt 2}(E+m)}\cr
              { p_{_{+}}^2/ 2(E+m) }\cr
               p_z\cr
                   {p_{_{+}}/{\sqrt 2}}\cr
                   0\cr}\,,\quad
u_{_{0}}( \vec p\,)\,=\,\pmatrix{{p_z p_{_{-}}/{\sqrt 2}(E+m)}\cr
                      m+\left[{p_{_{+}} p_{_{-}}/(E+m) }\right]\cr
                       -{p_z p_{_{+}}/{\sqrt 2}(E+m)}\cr
                       {p_{_{-}}/{\sqrt 2}}\cr
                          0\cr
                        {p_{_{+}}/{\sqrt 2}}\cr}\quad,\nonumber\\
u_{_{-1}}( \vec p\,)&\,=\,&\pmatrix{ { p_{_{-}}^2/ 2(E+m) }\cr
                             -{p_z p_{_{-}}/{\sqrt 2}(E+m)}\cr
                   m+\left[{(2p_z^2\,+\,p_{_{+}} p_{_{-}})/ 2(E+m)}\right]\cr
                      0\cr
                      {p_{_{-}}/{\sqrt 2}}\cr
                   -p_z\cr}\,,\quad\qquad
v_\sigma(\vec p\,) \,=\,
\left(
\begin{array}{ccc}
0 & {\quad}I \\
I & {\quad}0
\end{array}\right)
\,u_\sigma(\vec p\,)\label{uv}\quad.
\end{eqnarray}
}
In the above equation, we have defined $p_\pm\,=\,p_x\,\pm\,i\,p_y$.

Once we obtain the $(j,0)\oplus(0,j)$ spinors, %
 we make use of
Weinberg's \cite{SW}
observation that the general form of a field operator is dictated to %
us by
arguments of Poincar\'e covariance without any explicit  reference to a wave
equation. The $(j,0)\oplus(0,j)$ field operator reads 
\begin{equation}
\Psi(x) \,= \,
\sum_{\sigma=+j}^{-j}
\int {d^3p\over (2\pi)^{3} } {1\over 2\,\omega_{\vec p}}
\Big[ u_\sigma(\vec p\,)\, a_\sigma(\vec p\, )\, e^{-i p\cdot x}
+  v_\sigma(\vec p\,) \,b^\dagger_\sigma(\vec p\,) \,e^{i p\cdot x} \Bigr ]
\quad,\label{fo}
\end{equation}
where $\omega_{\vec p\,} = \sqrt{m^2 + {\vec p\,}^2}$; and
$[a_\sigma(\vec p\,),\,a^\dagger_{\sigma'}(\vec p^{\,\prime}]_\pm=
\,\delta_{\sigma\sigma'}
\delta^3(\vec p-\vec p^{\,\prime})$.

Since so far no wave equation  has been invoked, it is important to see how one
may obtain a wave equation.
To derive the wave equation satisfied by the
$(j,0)\oplus(0,j)$ spinors, we
observe that the general structure of the rest spinors given by Eq.
(\ref{rs}) implies \cite{CB} that (Burgard relations)
$
\phi_R(\vec 0\,)\, \,=\,\wp_{u,v}\,\phi_L(\vec 0\,)\,,
$
with
$
\wp_{u,v}= +1$ for the $u$-spinors and
$
\wp_{u,v}= -1$ for the $v$-spinors.
It may be parenthetically noted that in a similar context for spin one half,
Ryder [4, p.44] assumes the validity of an equation that reads,
$
\phi_R(\vec 0\,)=\phi_L(\vec 0\,)\,,
$
on the grounds that
``when a particle is at rest, one cannot define its spin as either left- or
right-handed.'' However, as we note, %
 this is simply a consequence of the general
structure of our theory --- moreover, in the process, we find an additional
minus sign (in the $\wp_{u,v}$ factor). This factor has
been found to have
profound significance for the internal consistency and consequences  in our
study.
 When
we couple the Burgard relations
with the transformation properties
(1) and (2), we [2f] obtain
$
\left[
\gamma_{\mu\nu\ldots\lambda}\,p^\mu p^\nu\ldots p^\lambda
\,-\,{\wp_{u,v}}\,m^{2j} I\right]\,\psi(\vec p\,)\,=\,0\,.
$
This equation, except for the  factor of ${\wp_{u,v}}$ attached to the
mass term, is identical to the Weinberg equation \cite{SW} for the
$(j,0)\oplus(0,j)$ spinors. The $2(2j+1)\times 2(2j+1)$-dimensional
$\gamma_{\mu\nu\ldots\lambda}$ matrices, with $2j$ Lorentz indices,
which appear here, can be found in Ref.
\cite{SW}, or in more explicit form in Ref. [2f].
For the $j=1/2$ case, this wave equation
 is found to be identical to the Dirac equation
in momentum space. For the $(1,0)\oplus(0,1)$
configuration-space free wave-functions 
$\psi(x)\equiv \psi(\vec p\,) \exp(-i\wp_{u,v}\,p\cdot x)$,
this wave equation  becomes
\footnotemark[3]
\footnotetext[3]
{Note that the $(1/2,0)\oplus(0,1/2)$ wave functions satisfy
$(i\gamma_\mu\partial^\mu\,-\,m\,I)\,\psi(x)\,=\,0$, {\it without} the
${\wp_{u,v}}$ attached to the mass term. It is readily seen that this
``cancellation'' of the ${\wp_{u,v}}$ when going from momentum space
to configuration space occurs for fermions, but not for bosons,
in general.}
\begin{equation}
\left(
\gamma_{\mu\nu}\partial^\mu\partial^\nu\,+\,
\wp_{u,v}\,m^2
\right)
\,\psi(x) \,=\,0\quad,\label{ea}
\end{equation}
with generalised canonical representation expressions for the $6\times 6$
ten
$\gamma_{\mu\nu}$-matrices given by
\begin{eqnarray}
\gamma_{00}\,=\,
\left(
\begin{array}{ccc}
I &{\,\,} &0\\
0 &{\,\,} &-I
\end{array}
\right),\quad
\gamma_{\ell0}\,=\,\gamma_{0\ell}\,=\,
\left(
\begin{array}{ccc}
0 &{\,\,}& -J_\ell\\
J_\ell&{\,\,} & 0
\end{array}
\right)\quad, \nonumber \\ \nonumber \\
\gamma_{\ell\jmath}\,=\,\gamma_{\jmath\ell}\,=\,
\left(
\begin{array}{ccc}
I & {\,\,}&0\\
0 &{\,\,}&-I
\end{array}
\right)
\,\eta_{\ell\jmath} \,+\,
\left(
\begin{array}{cc}
\{J_\ell,J_\jmath\} & 0\\
0 & -\,\{J_\ell,J_\jmath\}
\end{array}
\right)\quad.
\label{gm}
\end{eqnarray}
Here, $\eta_{\mu\nu}$ is the
flat space-time metric with $diag\,(1,\,-1,\,-1\,-1)$; and
$\ell,\,\jmath$
run over the spacial indices $1, \,2, \,3$.

Equation (\ref{ea}) has three $u_\sigma(\vec p\,)$ and
three $v_\sigma(\vec p\,)$ solutions given by Eq. (\ref{uv}). However
these solutions can also be interpreted as associated with not only
Einsteinian
$E=\pm\sqrt{{\vec p}^{\,2}+m^2}$ but also with tachyonic [2e,2f]
$E=\pm\sqrt{{\vec p}^{\,2}-m^2}$. This can be readily inferred
[2e,2f]
by studying the $12th$ order polynomial in $E$:
$Det \,\left(
\gamma_{\mu\nu}p^\mu p^\nu\,+\,
\wp_{u,v}\,m^2\,I
\right)=0\,$.
The tachyonic solutions are at this stage unphysical and can be ignored as long
as interactions are introduced in such a manner that they  do not induce
transitions between the  physical and unphysical solutions. In this context,
for the spin-one spinors, we introduce
\begin{equation}
P_u\,=\,{1\over {m^2}}\sum_{\sigma=0,\pm 1} u_\sigma(\vec p\,)\,
\overline {u}_\sigma(\vec p\,)\,,\quad
P_v\,=\,-\,{1\over {m^2}}\sum_{\sigma=0,\pm 1} v_\sigma(\vec p\,)\,
\overline {v}_\sigma(\vec p\,)\quad,\label{pupv}
\end{equation}
and verify that $P_u + P_v =\openone$,
 $P_u^2=P_u$, $P_v^2=P_v$, and $P_u\,P_v=0\,$.

In order to establish that the field operator defined by Eq. (\ref{fo})
describes a quantum field theory of the BWW type, we now show
that bosons and antibosons, within the $(j,0)\oplus(0,j)$ framework
developed above,  indeed have opposite intrinsic parity  and
well-defined $C$ and $T$ characteristics.
We begin with the classical considerations similar to the
ones found for the $(1/2,0)\oplus(0,1/2)$ Dirac field in the standard
texts, such as Ref. \cite{ON}. As the simplest example case we study the
$(1,0)\oplus(0,1)$ field in detail. As such,
 we seek a
parity-transformed wave function
$
\psi'(t',\,{\vec x}\,') \,=\,S(P)\,\psi(t,\,\vec x)\,;\quad
{x'}^\mu \,=\,{\left[\Lambda_P\right]^\mu}_\nu\,x^\nu\,
$
 [here: $\Lambda_P=diag\,(1,\,-1,\,-1,\,-1)$ so that $t'=t$ and
$\vec x^{\,\prime}=-\vec x$],
such that Eq. (\ref{ea}) holds true for $\psi^\prime(t',\,{\vec x}\,')$.
That is:
$
\left(\gamma_{\mu\nu}\,\partial^{\,\prime\mu}\,\partial^{\,\prime\nu}
 \,+\,\wp_{u,v}\,m^2 I\right)\,
\psi^\prime(t',\,{\vec x}\,'\,)\,=\,0\,.
$
It is a straightforward algebraic exercise to find that $S(P)$ must
simultaneously satisfy the following requirements
\begin{eqnarray}
[S(P)]^{-1}\,\gamma_{00}\,S(P)\,=\,\gamma_{00},
\quad
[S(P)]^{-1}\,\gamma_{0\jmath}\,S(P)\,=\,-\,
\gamma_{0\jmath}\quad,\nonumber \\
\,[S(P)]^{-1}\,\gamma_{\jmath0}\,S(P)\,=\,-\,
\gamma_{\jmath0},
\quad
[S(P)]^{-1}\,\gamma_{\ell\jmath}\,S(P)\,=\,\gamma_{\ell\jmath}
\quad.
\end{eqnarray}
Referring to Eqs. (\ref{gm}), we now note that while $\gamma_{00}$
commutes  with $\gamma_{\ell \jmath}$, it
anticommutes with $\gamma_{0\jmath}$
\begin{equation}
\left[\gamma_{00},\,\gamma_{\ell\jmath}\right]\,=\,
\left[\gamma_{00},\,\gamma_{\jmath\ell}\right]\,=\,0,\quad
\left\{\gamma_{00},\,\gamma_{0\jmath}\right\}\,=\,
\left\{\gamma_{00},\,\gamma_{\jmath0}\right\}\,=\,0\quad.
\end{equation}
As a result, confining to the norm-preserving transformations (and ignoring a
possible {\it global} phase factor
\footnotemark[4]
\footnotetext[4]{Such a global phase factor acquires crucial  importance for
constructing an %
internally consistent theory of Majorana-like $(j,0)\oplus(0,j)$
 fields.}),
we  identify $S(P)$ with
$\gamma_{00}$, yielding:
$
\psi'(t',\,{\vec x}\,') \,=\,\gamma_{00}\,\psi(t,\,\vec x)\quad
\Longleftrightarrow\quad
\psi'(t',\,{\vec x}\,') \,=\,\gamma_{00}\,\psi(t',\,-\,{\vec x}\,')
\,.
$
This prepares us to proceed to the field theoretic considerations.
The $(1,0)\oplus(0,1)$ matter field operator is defined by letting
$\sigma=0,\pm 1$ in the general expression (\ref{fo}).
The transformation properties of  the states
$
\vert\vec p,\,\sigma\rangle^u\,=\,a^\dagger_\sigma(\vec p\,)\,\vert\mbox{vac}
\rangle$ and
$
\vert\vec p,\,\sigma\rangle^v\,=\,b^\dagger_\sigma(\vec p\,)\,\vert\mbox{vac}
\rangle\,
$
are obtained from the condition
\begin{equation}
U(P)\,\Psi(t',\,\vec x\,'\,)\,[U(P)]^{-1}\,=\,
\gamma_{00}\,\Psi(t',\,-\,\vec x\,'\,)\label{uuin}
\quad,
\end{equation}
where $U(P)$ represents a unitary operator that governs the operation
of parity in the Fock space.
Using the definition of $\gamma_{00}$, Eqs. (\ref{gm}), and the
explicit expressions for the
$(1,0)\oplus(0,1)$ spinors $u_\sigma(\vec p\,)$ and $v_\sigma(\vec p\,)$
given by Eqs. (\ref{uv}), we find
\begin{equation}
\gamma_{00}\, u_\sigma(p')\,=\,+\,u_\sigma(p)\,,\quad
\gamma_{00}\, v_\sigma(p')\,=\,-\,v_\sigma(p)\quad,
\label{pt}
\end{equation}
with $p'$  the parity-transformed $p$  --- i.e.,
for $p^\mu\,=\,(p^0,\,\vec p\,)$,
$p^{\prime\mu}\,=\,(p^0,\,-\vec p\,)$. The observation (\ref{pt})
when coupled with the requirement (\ref{uuin}) immediately yields the
transformation properties of the
particle-antiparticle
 creation operators:
\begin{equation}
U(P)\,a^\dagger_\sigma(\vec p\,)\, [U(P)]^{-1}
\,=\,+\,a^\dagger_\sigma(-\,\vec p\,)\,,\,\,
U(P)\,b^\dagger_\sigma(\vec p\,)\, [U(P)]^{-1}
\,=\,-\,b^\dagger_\sigma(-\,\vec p\,)\,.
\end{equation}
Under the assumption that the vacuum is invariant under the parity
transformation, $U(P)\,\vert\mbox{vac} \rangle\,=\,\vert\mbox{vac} \rangle$, we
arrive at the result that the
\footnotemark[5]
\footnotetext[5]{The fuller justification for the terminology
``particle'' and ``antiparticle,''
apart from the convention  of what we call ``particle,''
will be realised when we consider the operation of $C$.}
 ``particles'' (described classically  by the
$u$-spinors) and ``antiparticles'' (described classically by the $v$-spinors)
have opposite relative intrinsic parities:
$
U(P)\,\vert\vec p,\,\sigma\rangle^u\,=\,+\,\vert\,-\vec p, \sigma
\rangle^u \,,\quad
U(P)\,\vert\vec p,\,\sigma\rangle^v\,=\,-\,\vert\,-\vec p, \sigma
\rangle^v \,.
$
This is precisely what we set out to prove. That is, the
$(1,0)\oplus(0,1)$ boson and antiboson  have opposite relative intrinsic
parity. As a consequence, the number of physical states, in comparison to the
description of a massive spin-one particle by the Proca vector potential
$A^\mu(x)$, is {\it doubled}  from $(2j+1)=3$ to $2(2j+1)=6$ in agreement
\footnotemark[6]
\footnotetext[6]{While our agreement with BWW \cite{EW} is complete, we
differ with Weinberg's claim
[4, footnote 13]   that the $(j,0)\oplus(0,j)$
fields have {\it same} relative intrinsic parity for bosons. The disagreement
with Ref. \cite{SW} arises because its author apparently %
did not realise that the
bosonic $v$-spinors are {\it not} solutions of his  proposed equation.
The factor $\wp_{u,v}$ in Eq. (\ref{ea})  {\it is required for internal
consistency} in the theory.}
with  BWW's work \cite{EW}.

Next we consider the operation of C. The charge conjugation operation C
must be carried through with a little greater care for bosons than
for fermions within the $(j,0)\oplus(0,j)$ framework developed here because of
the
$\wp_{u,v}$ factor in the mass term. For the $(1,0)\oplus(0,1)$ case,
at the classical level, we want
\begin{equation}
C:\quad
\left(\gamma_{\mu\nu}\,D^\mu_{(+)}\,D^\nu_{(+)}\,+\,m^2\right)\,u(x)\,=\,0
\,\,\,\longrightarrow\,\,\,
\left(\gamma_{\mu\nu}\,D^\mu_{(-)}\,D^\nu_{(-)}\,-\,m^2\right)\,v(x)\,=\,0
\quad,\label{c}
\end{equation}
where the local $U(1)$ gauge covariant derivatives are defined as:
$
D^\mu_{(+)}\,=\,\partial^\mu\,+\,i\,q\, A^\mu(x)\,,\quad
D^\mu_{(-)}\,=\,\partial^\mu\,-\,i\,q\, A^\mu(x)\,.
$
Again, a straightforward algebraic exercise yields the
result that for (\ref{c}) to occur we must have:
$
\psi(t,\vec x) \,\,\,\rightarrow \,\,\,S(C)\,\psi^\ast(t,\vec x)\,;
$
where $S(C)$ satisfies,
$
S(C)\,\gamma^\ast_{\mu\nu}\,[S(C)]^{-1}\,=\,-\,\gamma_{\mu\nu}\,.
$
We find that
$
S(C)\,=\,\eta_\sigma\,A\,\gamma_{00}\,,
$
with
\begin{equation}
A\,=\,\left(
\begin{array}{ccc}
0&{\,\,}&\Theta_{[1]}\\
\Theta_{[1]}&{\,\,}&0
\end{array}
\right)\,,\quad
\Theta_{[1]}\,=\,
\left(
\begin{array}{ccccc}
0&{\,\,}&0&{\,\,}&1\\
0&{\,\,}&-1&{\,\,}&0\\
1&{\,\,}&0&{\,\,}&0
\end{array}
\right)
\quad,
\end{equation}
and for convenience
we choose $\eta_{\pm 1}=+1$ and $\eta_{0}=-1$; and $\Theta_{[1]}$ is Wigner's
time-reversal operator \cite{TO} for spin-$1$. The effect of
charge conjugation in the Fock space is now immediately obtained
by using the easily verifiable identities:
$
S(C)\,u^\ast_{+1}(\vec p\,)\,=\,v_{-1}(\vec p\,)\,,\quad
S(C)\,u^\ast_{0}(\vec p\,)\,=\,v_{0}(\vec p\,)\,,\quad
S(C)\,u^\ast_{-1}(\vec p\,)\,=\,v_{+1}(\vec p\,)\,;
$
and the requirement
$
U(C)\,\Psi(x)\,[U(C)]^{-1}\,=\,\Psi^c(x)\,,
$
where
\begin{equation}
\Psi^c(x)\,=\,
\sum_{\sigma=0,\pm 1}
\int {d^3p\over (2\pi)^{3} } {1\over 2\,\omega_{\vec p}}
\Big[ S(C)\,u^\ast_\sigma(\vec p\,)\, a^\dagger_\sigma(\vec p\, )\,
e^{i p\cdot x}
+  S(C)\,v^\ast_\sigma(\vec p\,)
\,b_\sigma(\vec p\,) \,e^{-i p\cdot x} \Bigr ]
\quad.\label{pc}
\end{equation}
These considerations yield:
$
U(C)\,a^\dagger_\sigma(\vec p\,)\,[U(C)]^{-1}\,=\,b^\dagger_\sigma(\vec p\,)\,,
\quad
U(C)\,b^\dagger_\sigma(\vec p\,)\,[U(C)]^{-1}\,=\,a^\dagger_\sigma(\vec p\,)\,.
$
We thus see that the definition of charge-conjugation operation
C as given by (\ref{c}) indeed yields the correct picture in the Fock space:
$
U(C)\,\vert\vec p,\sigma\rangle^u\,=\,\vert\vec p,\sigma\rangle^v
$ and
$
U(C)\,\vert\vec p,\sigma\rangle^v\,=\,\vert\vec p,\sigma\rangle^u\,.
$

Finally, following Nachtmann \cite{ON}, we define the operation of time
reversal as a {\it product}
\footnotemark[7]
\footnotetext[7]{The {\it order} of the operations in the product that follows
is not important because the two operations are found to anticommute, and
therefore the ambiguity  of ordering only involves  an overall global phase
factor.}
of an operation, $S^\prime(\Lambda_T)$,
$
\psi(t,\vec x\,)\,\,\longrightarrow \,\,S^\prime(T)\,\psi(-t,\vec x\,),
$
which preserves the
form of Eq. (\ref{ea}) under $x^\mu\rightarrow {[\Lambda_T]^\mu}_\nu\,
x^{\,\prime\,\nu}$, $\Lambda_T=diag\,(-1,\,1,\,1,\,1)$,  {\it and}, based on
St\"uckelberg-Feynman \cite{SF} arguments, the operation of charge conjugation.
So, classically, under $T$ we have:
$
\quad\psi(t,\vec x)\,\,\longrightarrow\,\,
\psi^\prime(t,\vec x) \,=\,S^\prime(\Lambda_T)\,S(C)\,
\psi^\ast(-t,\vec x\,)\,.
$
We find that $S(T) \,\equiv\,S^\prime(\Lambda_T)\,S(C)$ is given by:
$
S(T)\,=\,(A\,\, global\,\, phase\,\, factor)\,\times\,\gamma_{00}\,
 \eta_\sigma\,A\,\gamma_{00}\,.
$
Taking note of the fact that $A$ anticommutes with $\gamma_{00}$,
$\{A,\,\gamma_{00}\}=0$, and dropping the resultant
global phase factor, we obtain $S(T)=\eta_\sigma\,A$.
In the Fock space, the %
above considerations are implemented by finding the effect of
an {\it anti-unitary} operator on the creation and annihilation operators via:
$
\left[V(T)\,\Psi(t,\vec x\,)\,[V(T)]^{-1}\right]^\dagger
\,=\,\Psi^\prime(t,\vec x\,)\,,
$
where
\begin{equation}
\Psi^{\prime}(t,\vec x\,)
\,=\,
\sum_{\sigma=0,\pm 1}
\int {d^3p\over (2\pi)^{3} } {1\over 2\,\omega_{\vec p}}
\Big[S(T)\, u^\ast_\sigma(\vec p\,)\, a^\dagger
_\sigma(\vec p\, )\, e^{i p\cdot x'}
+ S(T)\,  v^\ast_
\sigma(\vec p\,) \,b_\sigma(\vec p\,) \,e^{-i p\cdot x'} \Bigr ]
\quad,\label{t}
\end{equation}
with $x^{\prime\,\mu}=(-t,\vec x\,)$. Exploiting  the identities
\begin{equation}
S(T)\,u_\sigma^\ast(\vec p\,)\,=\,u_{-\sigma}(-\vec p\,)\,,\quad
S(T)\,v_\sigma^\ast(\vec p\,)\,=\,v_{-\sigma}(-\vec p\,)\,,
\end{equation}
we arrive at the result:
$
V(T)\,a^\dagger_\sigma(\vec p\,)\,[V(T)]^{-1}\,=\,
a^\dagger_{-\sigma}(-\vec p\,)\,,\quad
V(T)\,b^\dagger_\sigma(\vec p\,)\,[V(T)]^{-1}\,=\,
b^\dagger_{-\sigma}(-\vec p\,)\,.
$ \\
Therefore, if the vacuum is invariant under T, the physical states transform as
$V(T)\,\vert\vec p,\,\sigma\rangle \,=\, \vert-\vec p,\,-\sigma\rangle$ and
observables ${\cal O}\,\rightarrow {\cal O}^\prime\,=\, \left[V(T)\,{\cal O}\,
[V(T)]^{-1}\right]^\dagger$.

We thus see that the $(1,0)\oplus(0,1)$ field theory constructed above is
indeed of BWW type. What is left, in view of the work of Lee and Wick
\cite{LW}, is to explicitly show that the $(j,0)\oplus(0,j)$ field operator
(\ref{fo}) for spin one describes a {\it nonlocal} theory (in the sense to
become obvious below). A straightforward algebraic exercise yields the result
that for the $(1,0)\oplus(0,1)$ field operator associated with massive
particles,
 \begin{eqnarray}
\Big[&&\Psi_\alpha(t,\,\vec x\,)\,,\,\,\overline{\Psi}_\beta(t,\,
\vec x^{\,\prime}\,)
\Big]\,=\,\nonumber \\
&&\left( {1\over {2\,\pi}}\right)^6
\int  {{d^3\vec p}\over{2\,E(\vec p\,)} }
\sum_{\sigma=0,\pm 1} \bigg(
u_\sigma(\vec p\,)\,{\overline u}_\sigma(\vec p\,)
+
v_\sigma(-\,\vec p\,)\,{\overline v}_\sigma(-\,\vec p\,)
\bigg)_{\alpha\beta}\,
e^{i\vec p\cdot(\vec x-\vec x^\prime)}\quad.\label{nl}
\end{eqnarray}

The  nonlocality is now immediately inferred.
Using the explicit forms (\ref{uv}) of $u_{0,\pm 1}(\vec p\,)$
and $v_{0,\pm 1}(\vec p\,)$, we find
\begin{equation}
\sum_{\sigma=0,\pm 1} \bigg(
u_\sigma(\vec p\,)\,{\overline u}_\sigma(\vec p\,)
+
v_\sigma(-\,\vec p\,)\,{\overline v}_\sigma(-\,\vec p\,)
\bigg)\,=\,
\left(
\begin{array}{cc}
{\cal M} & 0\\
0 & -{\cal M}
\end{array}\right)
\quad,
\end{equation}
with
\begin{equation}
{\cal M}\,=\,
\left(
\begin{array}{ccccc}
E^2\,+\,p_z^2 &{\;\;}& \sqrt{2} \,p_-\, p_z&{\;\;} & p_-^2 \\
\sqrt{2}\, p_+\, p_z &{\;\;}& E^2\,+\,p_-\,p_+\, -\,p_z^2& {\;\;}&
-\,\sqrt{2}\, p_-\, p_z \\
p_+^2 &{\;\;}& -\,\sqrt{2} \,p_+\, p_z&{\;\;} & E^2\,+\,p_z^2
\end{array}
\right)\label{m}\quad.
\end{equation}
In Eq. (\ref{m}), $p_\pm=p_x\pm i p_y$; consequently, \\
$\left[\Psi_\alpha(t,\,\vec x\,)\,,\,\,\overline{\Psi}_\beta(t,\,\vec
x^\prime\,) \right]\,\ne\, (const.)\times\delta^3(\vec x-\vec x^\prime\,)$. \\
For comparison, we note that a
 similar calculation for the spin half Dirac case yields \\
$\left\{\Psi_\alpha(t,\,\vec x\,)\,,\,\,\overline{\Psi}_\beta(t,\,\vec
x^\prime\,) \right\}\,=\,(const.)\times
 \gamma^0_{\alpha\beta}\,\delta^3(\vec x-\vec x^{\,\prime}\,)\,$. \\
The crucial
property of the $(1/2,0)\oplus(0,1/2)$ representation space that enters
in obtaining this result is \\ %
$
\sum_{\sigma=\pm {1\over 2}}\Big(
u_\sigma(\vec p\,)\,{\overline u}_\sigma(\vec p\,)
\,+\,
v_\sigma(-\,\vec p\,)\,{\overline v}_\sigma(-\,\vec p\,)
\Big)_{\alpha\beta} \sim \left(\gamma^0\right)_{\alpha\beta}\,\,p_0
\,.
$
No corresponding {\it overall} factor of $E$ appears in
$\cal M$. If it did, the $E(\vec p\,)^{-1}$ factor in the
invariant element of phase space could be cancelled, leading to a
$\delta^3(\vec x-\vec x^\prime\,)$ and its derivatives
(Schwinger terms \cite{JS}) on the {\it rhs} of Eq. (\ref{nl}),
thus restoring  the locality.
For an alternate derivation of the nonlocality, the reader may
wish to refer to Ref. [2b]. For the sake of completeness, we note that
the momentum conjugate to the field operator $\Psi(x)$ for spin one
is given by $\Pi_0(x)=\overline{\Psi}(x)\,\gamma_{\mu 0}\,\partial^\mu$, and
$
\left[\Psi_\alpha(t,\,\vec x\,)\,,\,\,(\Pi_0)_\beta (t,\,\vec x^{\,\prime})
\right]\,=\,
-\,\delta^3(\vec x\,-\,\vec x^{\,\prime})\,
\partial^\mu\Psi_\alpha(t,\,\vec x\,)\,
\overline{\Psi}_\xi(t,\,\vec x\,)\,(\gamma_{\mu 0})_{\xi\beta}
\,.$ The physical interpretation of this last result requires further study.

A referee of one of our publications brought to our attention the fact %
that for the  $C$
and $P$ operators defined in the manuscript, $\{C,\,P\}=0$. As a
consequence, the particles and antiparticles in the $(1,0)\oplus(0,1)$
representation space do indeed have opposite intrinsic parity and  our
conclusion that we have constructed a Bargmann-Wightman-Wigner-type quantum
field theory is independent of any choice/conventions of phase factors.

To summarise, we note the $(j,0)\oplus(0,j)$ representation space associated
with massive particles is a concrete realisation of a quantum field theory,
envisaged many years ago by Bargmann, Wightman, and Wigner, in which bosons and
antibosons have opposite relative intrinsic parities. It is  our hope that our
detailed analysis of the $(j,0)\oplus(0,j)$ representation space would
supplement the canonical Bargmann-Wigner/Rarita-Schwinger \cite{BWRS} formalism
(where a boson and antiboson have same intrinsic parity) and open new
experimentally observable possibilities for the Poincar\'e covariant aspects of
hadrons and their propagation in nuclei.

\acknowledgements
D.V.A. extends his affectionate thanks to Christoph Burgard and Zimpoic V. for
our continuing conversations on this subject. It is also his pleasure to extend
his warmest thanks to the organisers of the conference. Particular thanks are
due to all the secondary and graduate school students who made his
 stay in Poland a truly warm personal experience. Cheers to them all !

\end{document}